\documentclass[trackchanges]{aastex7}
\received{April 18, 2025}
\accepted{\today}

\begin{document}

\title{\texttt{easyspec}: An open-source Python package for long-slit spectroscopy}

\author[orcid=0000-0001-5489-4925, sname='de Menezes']{Raniere de Menezes}
\affiliation{Centro Brasileiro de Pesquisas F\'isicas, 22290-180 Rio de Janeiro, RJ, Brazil}
\email[show]{raniere@cbpf.br}  

\author[orcid=0000-0002-1704-9850, sname='Massaro']{Francesco Massaro} 
\affiliation{Dipartimento di Fisica, Universit\`a degli Studi di Torino, via Pietro Giuria 1, I-10125 Turin, Italy}
\affiliation{Istituto Nazionale di Fisica Nucleare, Sezione di Torino, via Pietro Giuria 1, I-10125 Turin, Italy}
\affiliation{INAF -- Osservatorio Astrofisico di Torino, via Osservatorio 20, 10025 Pino Torinese, Italy}
\affiliation{Consorzio Interuniversitario per la Fisica Spaziale (CIFS), via Pietro Giuria 1, 10125 Turin, Italy}
\email{f.massaro@unito.it}

\author[orcid=0000-0002-6548-5622, sname=Negro]{Michela Negro}
\affiliation{Department of Physics \& Astronomy, Louisiana State University, Baton Rouge, LA 70803, USA}
\email{fakeemail3@google.com}

\author[orcid=0000-0003-1784-2784, sname=Raiteri]{Claudia M. Raiteri}
\affiliation{INAF -- Osservatorio Astrofisico di Torino, via Osservatorio 20, 10025 Pino Torinese, Italy}
\email{claudia.raiteri@inaf.it}

\author[orcid=0000-0003-0032-9538,sname='Pe\~na-Herazo']{Harold Pe\~na-Herazo}
\affiliation{East Asian Observatory, 660 N. A'ohoku Place, Hilo, HI 96720, USA}
\email{h.pena@eaobservatory.org}

\author[orcid=0000-0002-0433-9656,sname='Acosta-Pulido']{Jose A. Acosta-Pulido}
\affiliation{Instituto de Astrof\'isica de Canarias, Calle V\'ia L\'actea, s/n, 38205 La Laguna, Tenerife, Spain}
\affiliation{Departamento de Astrof\'isica, Universidad de La Laguna, 38206 La Laguna, Tenerife, Spain}
\email{jose.acosta@iac.es}

\begin{abstract}

In modern-day astronomy, near-infrared, optical, and ultraviolet spectroscopy are indispensable for studying a wide range of phenomena, from measuring black hole masses to analyzing chemical abundances in stellar atmospheres. However, spectroscopic data reduction is often performed using instrument-specific pipelines or legacy software well-established and robust within the community that are often challenging to implement and script in modern astrophysical workflows. In this work, we introduce easyspec, a new Python package designed for long-slit spectroscopy, capable of reducing, extracting, and analyzing spectra from a wide range of instruments—provided they deliver raw FITS files, the standard format for most optical telescopes worldwide. This package is built upon the well-established long-slit spectroscopy routines of the Image Reduction and Analysis Facility (IRAF), integrating modern coding techniques and advanced fitting algorithms based on Markov Chain Monte Carlo (MCMC) simulations. We present a user-friendly open-source Python package that can be easily incorporated into customized pipelines for more complex analyses. To validate its capabilities, we apply easyspec to the active galactic nucleus G4Jy 1709, observed with the DOLORES spectrograph at the Telescopio Nazionale Galileo, measuring its redshift and estimating its supermassive black hole mass. Finally, we compare our results with a previous IRAF-based study.

\end{abstract}



\section{Introduction}
\label{sec:intro}

Long-slit spectroscopy is one of the most powerful diagnostic tools in astronomy. In this technique, light from an astrophysical object--such as a galaxy, nebula, or star--passes through a narrow, elongated slit before being dispersed by a diffraction grating or prism, breaking it into its component wavelengths. A charge-coupled device (CCD) or a complementary metal–oxide–semiconductor (CMOS) detector then records the dispersed light, producing a two-dimensional spectrum where one axis represents wavelength and the other, spatial position along the slit.

Raw data from astronomical cameras are often contaminated by various noise sources, including sensor bias, thermal noise, detector sensitivity variations, and cosmic-ray hits. To ensure accurate spectral analysis, these artifacts must be carefully removed before extracting and calibrating the spectrum in both wavelength and flux. Traditionally, the Image Reduction and Analysis Facility \citep[IRAF;][]{tody1986iraf,pena2001pyraf} has been the standard tool for these tasks and remains widely used in the community \citep[e.g.,][]{sen2025discovery,guerrero2025confirmation,schonell2025jet}\footnote{Support for IRAF was discontinued in 2013 by the NOAO, and it is currently maintained by community effort \url{https://iraf-community.github.io}.}.

Other legacy software, such as the STARLINK library\footnote{Officially discontinued in 2005 and since then being maintained by the East Asian Observatory: \url{https://starlink.eao.hawaii.edu/starlink}} \citep{currie2014starlink,berry2022starlink2}, provides valuable tools for astronomical data reduction. However, with the rise of Python as a leading high-level language for astronomical analysis, some legacy routines have been integrated into Astropy \citep{astropy2013A&A...558A..33A,astropy2018AJ....156..123A}. Additionally, new software for astronomical data reduction has been actively developed \citep[e.g.,][]{pypeit2020joss_pub,lam2023aspired,pickering2024specreduce}, offering modern alternatives to traditional tools.

In this work, we introduce easyspec, a user-friendly, open-source Python package for data reduction, spectral extraction, and line fitting in long-slit spectroscopy of astrophysical objects. In the context of modern spectroscopy tools, easyspec distinguishes itself by integrating both data reduction and spectral analysis within a single package. It places particular emphasis on robust line-fitting capabilities and a highly visual, interactive workflow. Diagnostic plots are provided at nearly every stage of the reduction and analysis processes, offering users intuitive insights and enhancing transparency throughout their analysis. To showcase its capabilities, we apply easyspec to the raw spectroscopic data of the active galactic nucleus (AGN) G4Jy 1709 \citep[previously studied by][]{holt2008fast,massaro2023powerful}, also known as PKS 2135-20, observed in September 2024 with the 3.58-meter Telescopio Nazionale Galileo (TNG) in the Canary Islands, Spain. We estimate the mass of its supermassive black hole and measure its redshift. By streamlining the data analysis process, easyspec enhances user experience and significantly reduces the time required for near-infrared, optical, and near-ultraviolet astronomical spectroscopy.

The tutorials for easyspec and detailed documentation are available on GitHub\footnote{\url{https://github.com/ranieremenezes/easyspec}} and the official documentation webpage\footnote{\url{https://easyspec.readthedocs.io/en/latest/index.html}}. Although here we focus on spectroscopic data observed with TNG, we have successfully tested easyspec in data from other telescopes, such as the 4.1m Southern Astrophysical Research Telescope (SOAR), at Cerro Pach\'on, Chile, and the 1.6m Perkin-Elmer telescope at the Observat\'orio do Pico dos Dias (OPD), Brazil. This paper is structured as follows: Sect. \ref{sec:installation} outlines the software's dependencies and provides an overview of its functionality. Sect. \ref{sec:application} describes its application to G4Jy 1709, covering data reduction, spectral extraction, and line fitting, along with a detailed explanation of the adopted methods. Sect. \ref{sec:results} presents our results on the estimation of the supermassive black hole mass, while Sect. \ref{sec:validation} discusses the validation of easyspec by comparing it with other software and previous studies. Finally, Sect. \ref{sec:discussion_and_conclusions} summarizes our findings and highlights the potential applications of easyspec.


\section{Code overview and main dependencies}
\label{sec:installation}

The current release of easyspec \citep[v1.0.0.9;][]{deMenezes2025easyspec_zenodo} is available on the Python Package Index\footnote{\url{https://pypi.org/project/easyspec/}} (PyPI) and GitHub, along with detailed installation and usage instructions. The package leverages several well-established Python tools for astronomy and data analysis, including Astropy \citep{astropy2013A&A...558A..33A,astropy2018AJ....156..123A} for handling FITS files, ccdproc \citep{matt_craig2017ccdproc} for gain correction and cosmic-ray removal \citep[through the L.A. Cosmic algorithm detailed in][]{van2001cosmic,curtis_mccully_2018lacosmic}, dust\_extinction \citep{gordon2024dust_extinction} for extinction corrections, emcee and corner \citep{foreman2013emcee,foreman2016corner} for Markov chain Monte Carlo (MCMC) line fitting, and the widely used packages Matplotlib, NumPy, and SciPy \citep{Hunter2007matplotlib,harris2020Numpy,Virtanen2020SciPy}.

The easyspec package is designed to process raw spectroscopic FITS files and is structured into three main modules: cleaning, spectral extraction, and line fitting. In the next section, we apply easyspec to the AGN G4Jy 1709 and provide a detailed explanation of each step in the data reduction and line-fitting process. For a more in-depth analysis, we refer readers to the tutorial notebooks available on GitHub\footnote{\url{https://github.com/ranieremenezes/easyspec/tree/main/tutorial}}, where we describe step-by-step what each easyspec function is doing and give general advice on how to do long-slit spectroscopy. 


\section{Application to the active galactic nucleus G4Jy 1709}
\label{sec:application}

In September 2024, we observed the AGN G4Jy 1709 using the low-resolution spectrograph DOLORES at the 3.58-meter Telescopio Nazionale Galileo, located at the Roque de los Muchachos Observatory in the Canary Islands, Spain. We performed three 1800-second exposures with a $1.0^{\prime\prime}$ slit, no filter, and the LR-R grism (300 lines/mm). For calibration, we acquired 15 bias frames (zero exposure, closed shutter), seven sky flats (0.9 s each), seven lamp\footnote{Reference lamp webpage: \url{https://www.tng.iac.es/instruments/lrs/}} exposures (Ar, Ne+Hg, Kr; 0.9 s each), and four exposures (5 s each) for the standard star BD+33d2642, using a slit width of $1.0^{\prime\prime}$ for all of them. The major noise sources and instrumental signal offsets affecting these data are:

\begin{figure}
    \centering
    \includegraphics[scale=0.6]{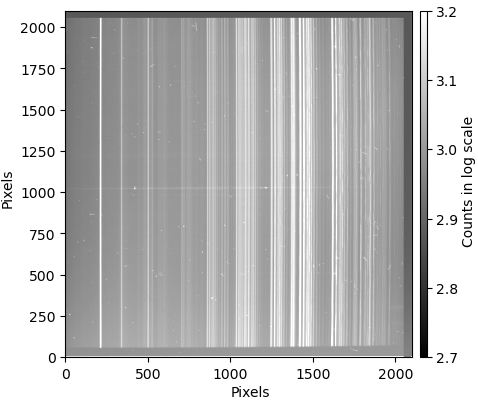}
    \caption{Raw spectrum of G4Jy 1709 for an exposure of 1800 seconds. The spectrum of our target is the dim horizontal line near the y-axis pixel 1000. In the edges of the image, we can see the dark patches in the light sensor where the light coming from the aperture does not arrive.}
    \label{fig:raw_spec}
\end{figure}

\begin{itemize}
    \item Sensor bias--An electronic offset introduced by the detector's readout electronics. It is measured by taking a zero-second exposure with the shutter closed and serves as one of the most fundamental calibration steps, as it affects all images. For CCDs, the bias includes a fixed direct current offset in the amplifier, as well as contributions from low-order structure and random noise introduced during the readout process. To correct for this effect, the bias must be subtracted from the data.
    \item Dark current--Thermal noise that can be mitigated by cooling the detector. This image is obtained in exposures taken with the shutter closed for the same duration as the target observation. To remove this noise, the dark frame must be subtracted from the data. For our observations of G4Jy 1709 this is not an issue.
    \item Flat-field variations--Pixel-to-pixel sensitivity differences in the light sensor. These variations can be measured by taking a relatively short exposure of a uniformly illuminated white screen inside the telescope dome (or using a sky-flat). After subtracting bias and dark frames (if needed), the 2D spectrum must be divided by the flat field to correct for these inhomogeneities. This correction must be applied to any image done with a non-zero exposure and open shutter (i.e., it does not affect the bias and dark frames). It is important to acquire flat-field frames with the highest possible signal-to-noise ratio in order to minimize the propagation of additional noise into the science data during the flat-field correction.
    \item Cosmic-ray strikes--Longer exposures accumulate more cosmic-ray hits. If several exposures (typically $\geq 5$) of the same target are available, cosmic rays can be removed by taking the median of all exposures (avoid using the average since it cannot get rid of the strongest cosmic-ray strikes). If only a few exposures are available, as in our case, specialized algorithms (see Sect. \ref{sec:data_cleaning}) can detect and remove cosmic rays by identifying their sharp edges.
\end{itemize}

In Figure \ref{fig:raw_spec} we show one of the three raw spectra. The x-axis corresponds to the dispersion axis, while the AGN spectrum appears as a faint horizontal line around the y-axis pixel 1000. At this stage, the image includes randomly scattered cosmic-ray hits. To start the data reduction, we import the easyspec \texttt{cleaning()} class and load all data with the function \texttt{cleaning.data\_paths()}. In this function, the user must specify the paths to the bias, flats, lamps, standard star, and target data, which will then be converted into a dictionary containing the name of each file as the dictionary keywords and the loaded data as the dictionary values. Notably, easyspec never modifies the original files, ensuring they remain intact in their original directories.

\subsection{Data cleaning}
\label{sec:data_cleaning}

\begin{figure}
    \centering
    \includegraphics[scale=0.7]{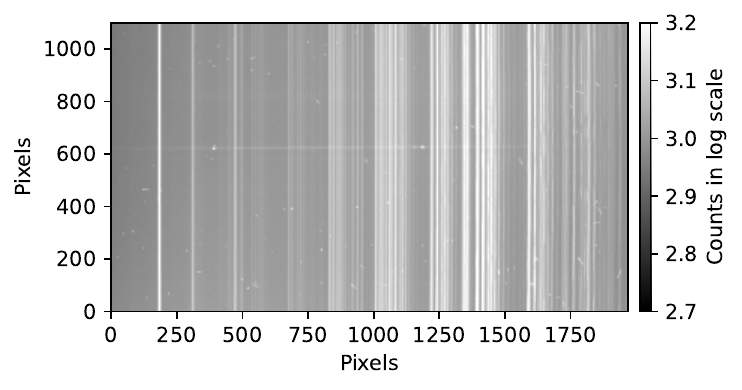}\\
    \includegraphics[scale=0.7]{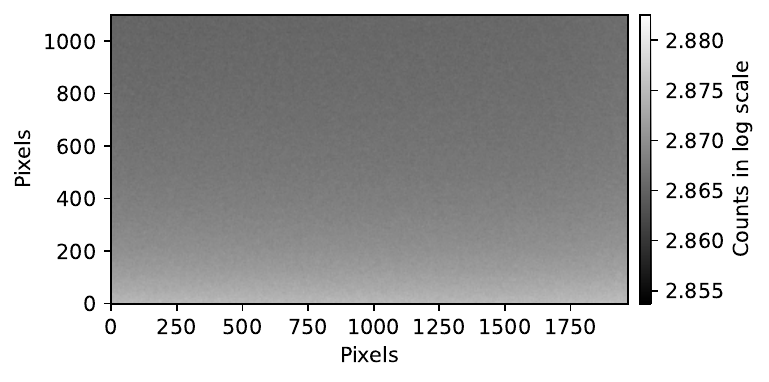}\\
    \includegraphics[scale=0.7]{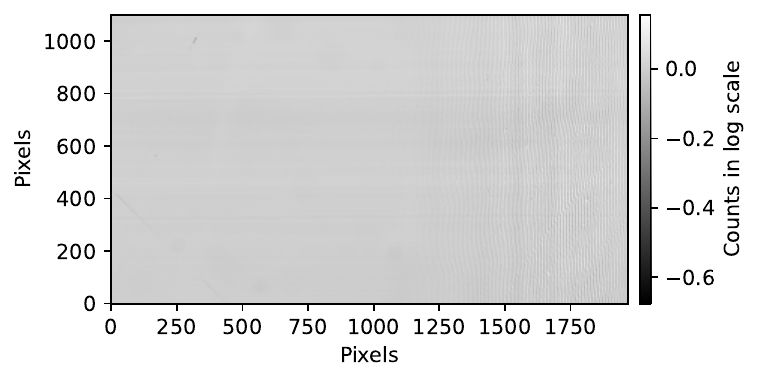}
    \caption{Trimmed raw spectrum (top), median bias (middle), and normalized median flat-field (bottom) for our observations.}
    \label{fig:masters}
\end{figure}

The first step in data cleaning is trimming the data. This step is typically used to remove the overscan region or areas of the image that did not capture light from the slit (see the edges of Figure \ref{fig:raw_spec}). We apply the easyspec function \texttt{cleaning.trim()} to crop all data files stored in the dictionary produced by \texttt{data\_paths()}, selecting the pixel range $x_1=30$, $x_2=2000$, $y_1=400$, $y_2=1500$. The output is a new dictionary containing the cropped data, with the same keys as the original. The top panel of Figure \ref{fig:masters} displays one of the trimmed spectra.

Next, we compute the average, median, or mode bias using the 15 bias exposures mentioned earlier. This step is easily performed with the easyspec function \texttt{cleaning.master($``bias"$)} applied to the subset of bias files in the trimmed data dictionary. In this case, we use the median bias, shown in the middle panel of Figure \ref{fig:masters}. Since all camera images are affected by bias, we subtract the median bias from all of them at once using the easyspec function \texttt{cleaning.debias()}.

Since we are using a refrigerated camera, thermal noise in the CCD is negligible\footnote{For DOLORES, dark current is negligible even for long exposures: \url{https://www.tng.iac.es/instruments/lrs/}}. However, if needed, the user can remove thermal noise using the easyspec function \texttt{cleaning.master($``dark"$)}\footnote{More details here: \url{https://github.com/ranieremenezes/easyspec/blob/main/tutorial/Image_cleaning_easyspec.ipynb}}. The next step is flat-field correction, as CCD sensitivity varies across pixels. We compute the median flat-field using the function \texttt{cleaning.master($``flat"$)}. However, since CCD sensitivity and grating diffraction efficiency depend on wavelength, the median flat-field often exhibits a strong intensity gradient along the dispersion axis.

\begin{figure}
    \centering
    \includegraphics[scale=0.7]{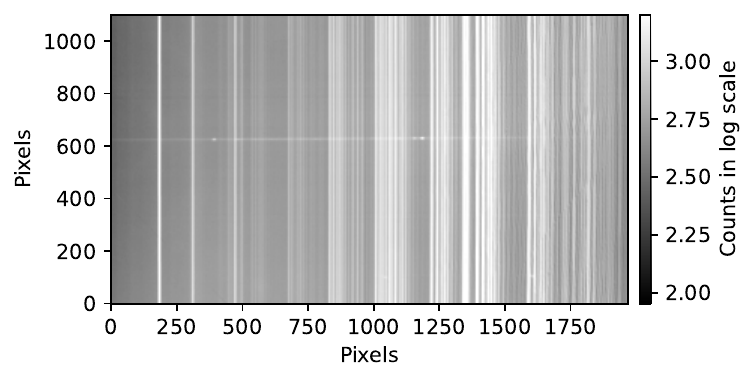}
    \caption{Cosmic-ray corrected and stacked 2D spectrum for G4Jy 1709. We can see the stacked spectrum as the horizontal line just above the y-axis pixel 600. The vertical lines are all sky lines, produced by excited molecules in Earth's atmosphere.}
    \label{fig:stacked}
\end{figure}

This is not the correction we seek—we are primarily interested in removing pixel-to-pixel sensitivity variations independent of wavelength. To achieve this, we normalize the median flat by fitting a polynomial to its intensity profile along the dispersion axis (if the flat-field presents wiggles, which are difficult to fit using polynomials, this function allows for the use of a median filter with a customizable smoothing window), using the function \texttt{cleaning.norm\_master\_flat()}. The resulting normalized median flat is shown in the bottom panel of Figure \ref{fig:masters}. Finally, we apply the normalized flat-field correction to all debiased data files with the function \texttt{cleaning.flatten()}.

With the camera's intrinsic features corrected, the next step is cosmic-ray removal. We use the easyspec function \texttt{CR\_and\_gain\_corrections()}, which adopts the LACosmic algorithm, originally developed for IRAF by \cite{van2001cosmic}. At this stage, we also correct for CCD gain and read noise, adopting a Laplacian-to-noise limit of $7\sigma$ for cosmic-ray detection. Finally, we stack the three cleaned spectra to produce a final 2D spectrum, shown in Figure \ref{fig:stacked}. One of easyspec's key features is its ability to apply this same process to lamp, standard star, and even other target spectra with minimal additional effort (see the GitHub tutorial for details).

\subsection{Spectral extraction}

\begin{figure}
    \centering
    \includegraphics[scale=0.7]{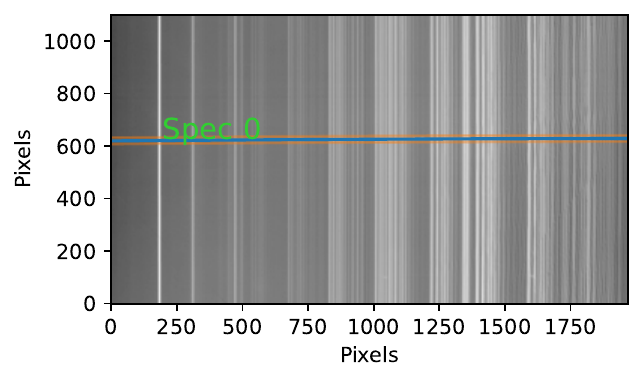}
    \caption{Spectral trace found with easyspec (blue line). The orange band around the trace is the region at which the spectrum will be extracted with a Gaussian.}
    \label{fig:trace}
\end{figure}

With the cleaned 2D spectrum in hand, the next step is to extract its trace (see Fig. \ref{fig:trace}). Although the dispersed target spectrum may appear as a straight horizontal line, this is rarely precisely the case. In most cases, it follows a shallow parabolic curve with minimal curvature, appearing almost straight. We perform this step using the easyspec class \texttt{extraction()}, specifically the function \texttt{extraction.tracing()}, which fits a polynomial to recover the trace of one or multiple spectra in an image. The tracing method can be set to:

\begin{itemize}
    \item argmax or moments: for extracting the strongest spectrum in the image.
    \item multi: for extracting multiple spectra from the cleaned image.
\end{itemize}
For our target, we use the argmax approach, which selects the pixel with the maximum value in each column and fits a polynomial to these points. This function also allows customization of the polynomial order, minimum average counts per spectrum, and distance between traces (either a constant or a user-defined list). The spectral trace found with easyspec is shown in Figure \ref{fig:trace} and is automatically assigned the name spec 0. If multiple spectra were present in the slit, the multi method would assign them names sequentially from bottom to top: spec 0, spec 1, spec 2, and so on. This method finds all local maxima by direct comparison with neighboring values using the \texttt{scipy} function \texttt{signal.find\_peaks()} applied to take the 1-D array of average x-axis image values.

\begin{figure*}
    \centering
    \includegraphics[width=\linewidth]{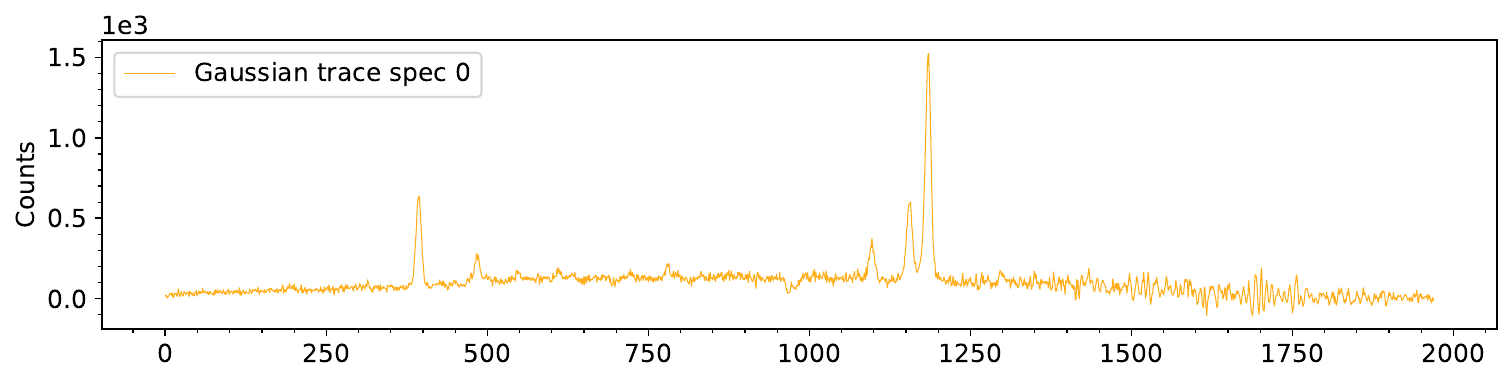}
    \includegraphics[width=\linewidth]{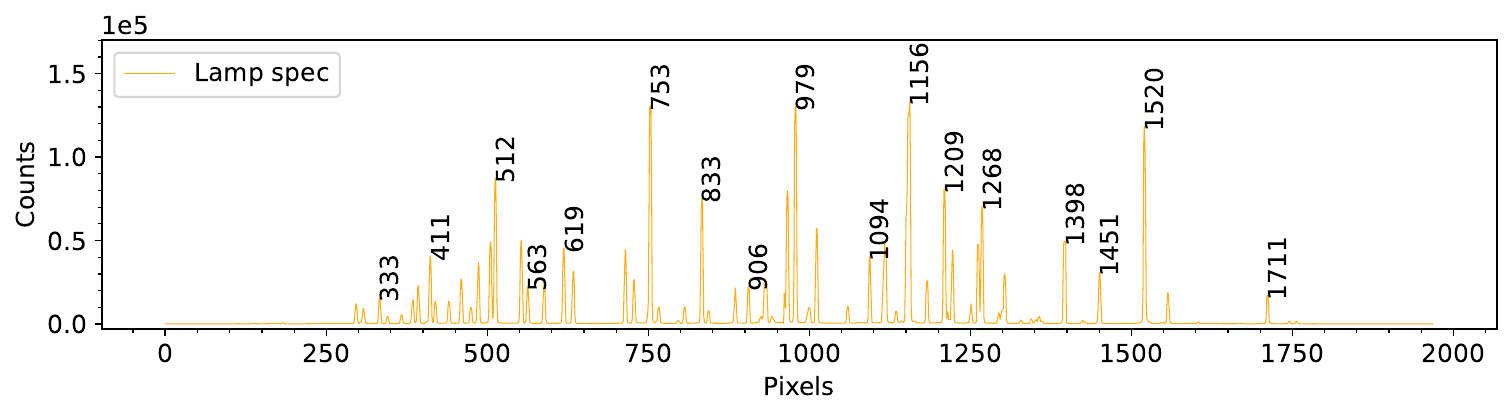}
    \caption{Non-calibrated spectra for G4Jy 1709 (top) and the Ar, Ne+Hg, Kr lamp (bottom). The numbers over the lines represent the pixel in the x-axis where the peak of each line is found. These lines are strategically selected by easyspec, such that they are relatively strong and far apart from each other.}
    \label{fig:non_calib_spec}
\end{figure*}

Once the trace is identified, we proceed with spectrum extraction using the \texttt{extraction.extracting()} function, which applies a Gaussian-weighted model to extract one or more spectra from the image. The initial guess for the Gaussian width is user-defined (with a default value of 3.5 pixels) and is subsequently refined by fitting it to the linearized trace profile. This profile is obtained by averaging the vertical dispersion (i.e., perpendicular to the wavelength axis) over the linearized trace. This function also estimates the systematic uncertainty of the extracted spectrum by performing a Monte Carlo simulation over the uncertainty ranges of the fitted Gaussian parameters. The default number of iterations is 50, though the user can specify any positive integer. The simulation provides an independent error estimate for each spectral bin, which is saved to a text file alongside the calibrated spectrum at the end of the analysis process. After extracting the target spectrum, we use the same trace to extract the lamp spectrum from the stacked 2D lamp image. To estimate the sky background, the trace is shifted by $\pm 30$ pixels along the y-axis (a user-defined value) to extract two sky spectra—one above and one below the original trace. The average of these sky spectra is then subtracted from the target spectrum. The resulting extracted target and lamp spectra are shown in Figure \ref{fig:non_calib_spec}. For those familiar with AGN spectroscopy, the Oxygen and Hydrogen emission lines in the G4Jy 1709 spectrum are immediately recognizable. However, at this stage, the spectrum remains uncalibrated, with the x-axis in pixels and the y-axis in raw counts. The detailed line identification process will be discussed in Sect. \ref{sec:line_fitting}.

\begin{figure}
    \centering
    \includegraphics[scale=0.7]{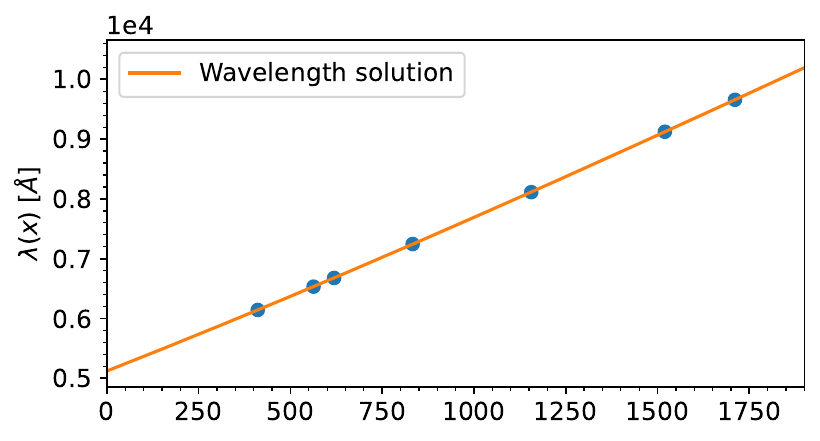}
    \includegraphics[scale=0.7]{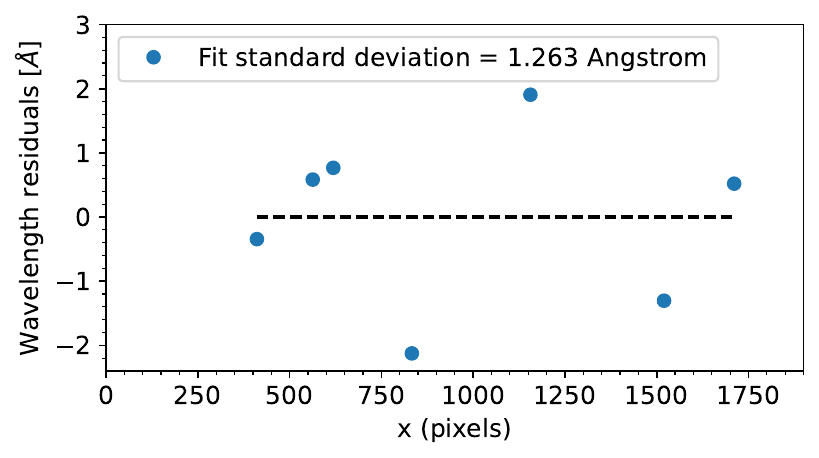}
    \caption{Wavelength solution (top) and residuals (bottom) for the spectrum of G4Jy 1709.}
    \label{fig:wavelength_solution}
\end{figure}

We now compare our lamp spectrum with the archival lamp spectrum available on the TNG website\footnote{\url{https://www.tng.iac.es/instruments/lrs/}} and use the easyspec function \texttt{extraction.wavelength\_calibration()} to determine the wavelength solution. This function takes an array of selected emission lines from the lamp spectrum (see bottom panel of Figure \ref{fig:non_calib_spec}) and fits an i-th order polynomial (here we choose third order), mapping wavelengths as a function of dispersion-axis pixels. Out of all the lamp lines identified with easyspec, we select those seven that have a corresponding value in the online list. The resulting wavelength solution and fit residuals are presented in Figure \ref{fig:wavelength_solution}. The standard deviation of the fit, $1.263\mathrm{\AA}$, is approximately half the average spectral resolution element, $2.623\mathrm{\AA}$, indicating a well-constrained calibration. This standard deviation is taken as the systematic error for the wavelength calibration and is saved alongside the calibrated spectrum at the end of the analysis.

\begin{figure*}
    \centering
    \includegraphics[width=\linewidth]{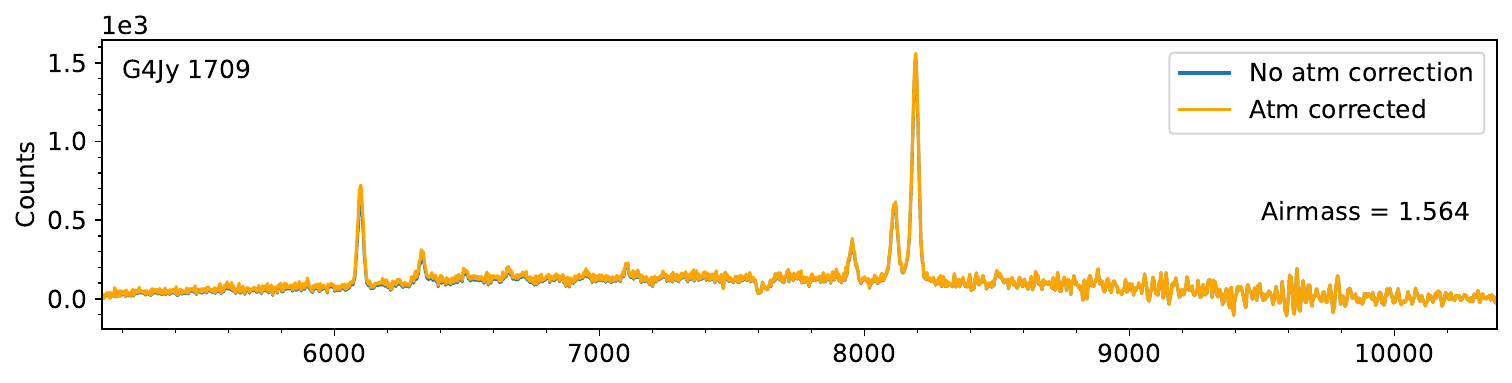}
    \includegraphics[width=\linewidth]{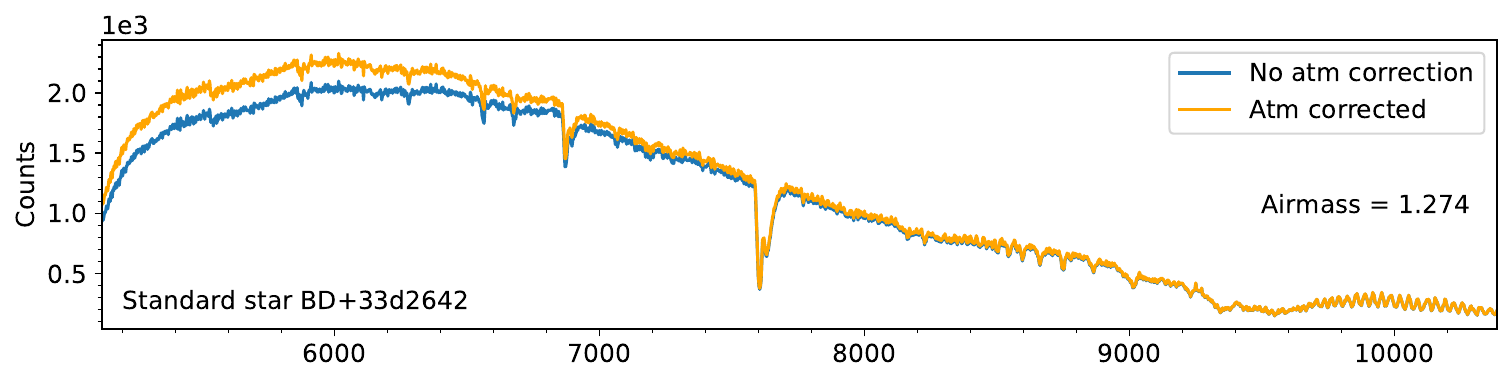}
    \includegraphics[width=\linewidth]{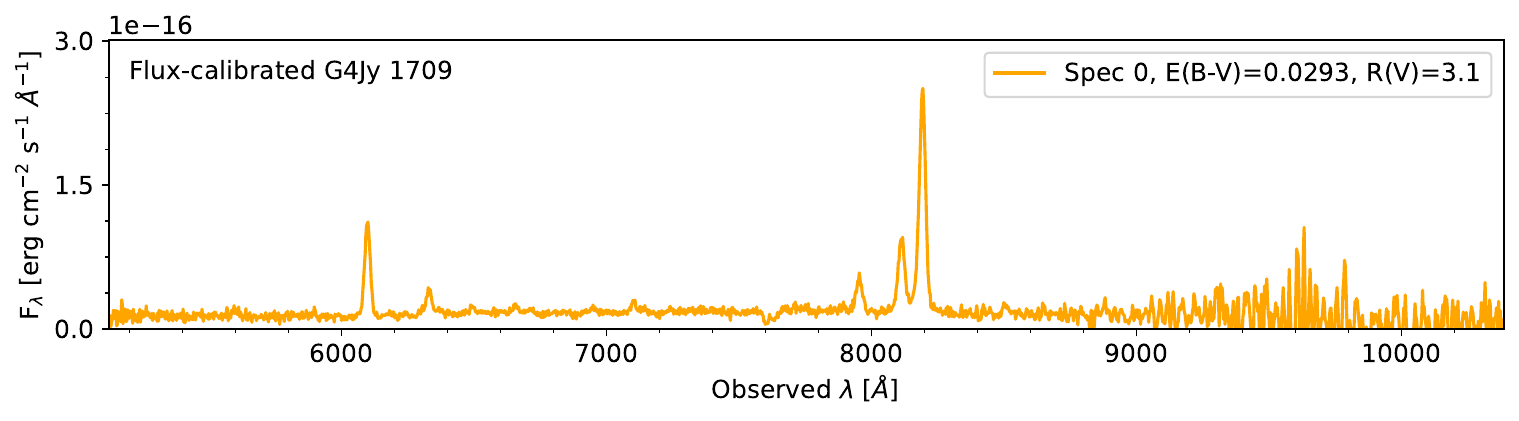}
    \caption{Target and standard star calibrated for atmospheric extinction (top and middle panels, respectively), and flux-calibrated target spectrum (bottom panel). The effect of atmospheric extinction is more relevant towards bluer wavelengths. We also see the presence of a relatively strong telluric line at $\lambda \approx 7600 \AA$.}
    \label{fig:final_spec}
\end{figure*}

With the wavelength solutions established individually for the target and standard star, we use the function \texttt{extraction.extinction\_correction()} to correct for the wavelength-dependent atmospheric extinction based on their respective air masses, using the extinction curve of Roque de los Muchachos\footnote{\url{https://www.ing.iac.es/Astronomy/observing/manuals/ps/tech_notes/tn031.pdf}}. These corrected spectra are displayed in the top and middle panels of Figure \ref{fig:final_spec}.

The final step is flux calibration. We derive the flux correction curve using the \texttt{std\_star\_normalization()} function, which compares the measured standard star spectrum with its archival reference spectrum. In this case, we use the BD+33d2642 standard star spectrum from the IRSCAL database. This function begins by extracting the continuum emission from both the archival and observed standard star spectra (already normalized by exposure time) using a median filter, with kernel sizes defined by the user (default is 101 for the observed spectrum and 11 for the archival spectrum). If undesired features--such as strong telluric absorption lines--affect the continuum estimation, the user can exclude specific wavelength regions from the median filtering by providing a list of intervals to ignore. The correction factor curve is then computed by dividing the interpolated archival continuum by the interpolated observed continuum. This flux calibration is subsequently applied to the exposure-corrected spectrum of G4Jy 1709 using the function \texttt{target\_flux\_calibration()}. Finally, we correct for Galactic reddening using the Galactic visual extinction estimate from \cite{schlafly2011measuring} and \cite{IRSA2022doi}\footnote{Extinction values for every direction in the sky can be found here: \url{https://irsa.ipac.caltech.edu/applications/DUST/}.}. This correction is needed because as light travels through the interstellar medium, dust particles in the Galaxy scatter and absorb shorter-wavelength photons more efficiently than longer-wavelength photons. This process causes the observed light to appear redder than its intrinsic color. The fully calibrated spectrum is presented in the bottom panel of Figure \ref{fig:final_spec} and saved in a text file together with the systematic uncertainties in wavelengths estimated as the standard deviation of the wavelength solution.

\subsection{Line fitting}
\label{sec:line_fitting}

\begin{figure*}
    \includegraphics[width=\linewidth]{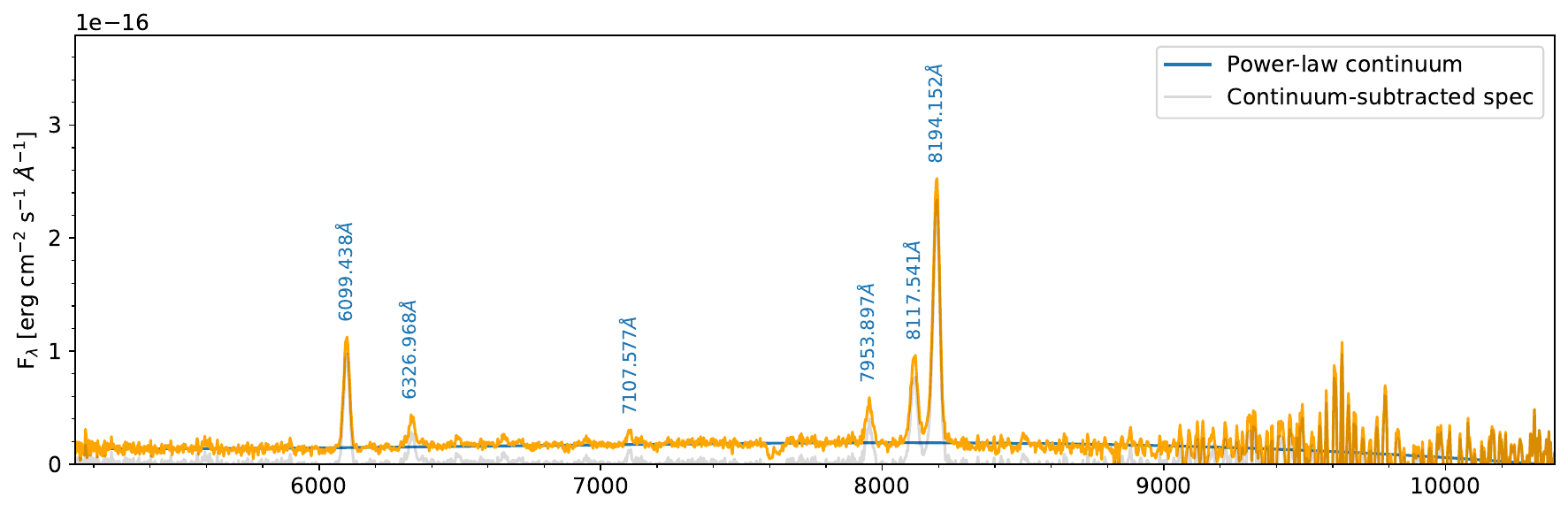}
    \includegraphics[width=\linewidth]{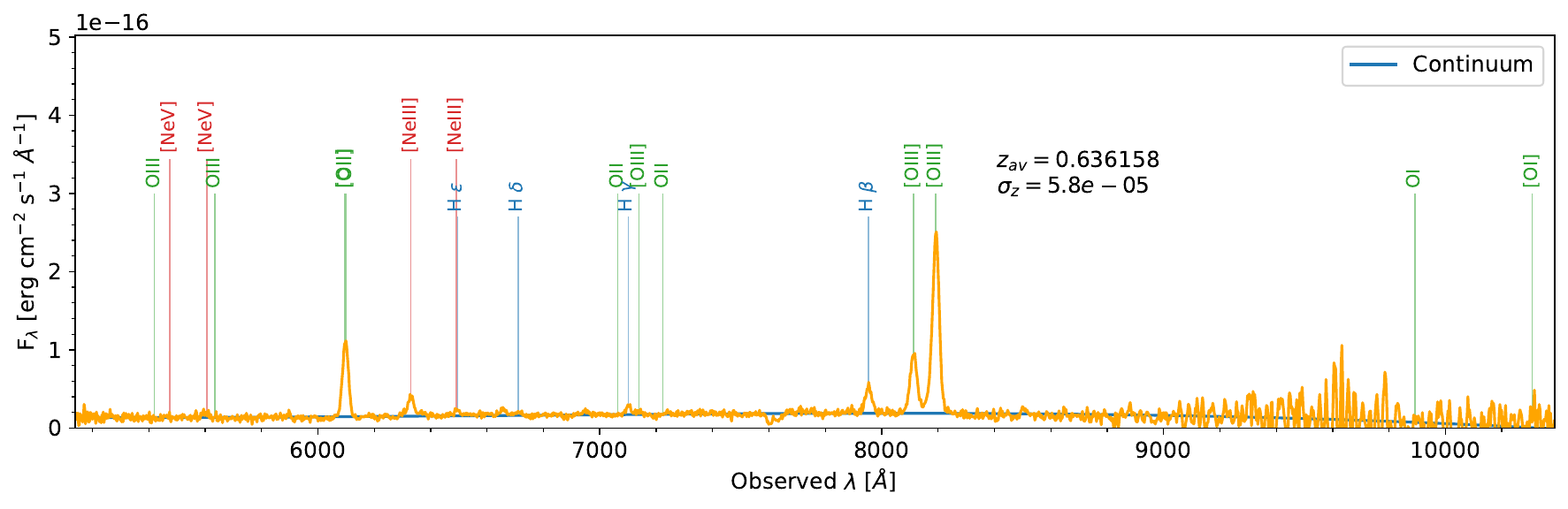}
    \caption{Detection (top) and assignment (bottom) of the lines done by easyspec. In the top panel, all lines with more than $5\sigma$ with respect to the local background are detected. In the bottom panel, the line fitting gives us the average redshift $z_{av}$ of the 6 lines analyzed and its standard deviation. In this panel, we also see the over-plotted Hydrogen, Oxygen, and Neon lines from the easyspec database. It is clear that some weak lines, like [NeV] and [OI] are lost below the adopted significance threshold.}
    \label{fig:lines}
\end{figure*}

With the calibrated spectrum ready, we now proceed to the easyspec class \texttt{analysis()}. The function \texttt{analysis.find\_lines()} takes user-defined continuum zones as input and then identifies emission and absorption lines with significances above a specified threshold--in this case, $5\sigma$--with respect to the closest user-defined continuum. Here, significance is defined as the line height divided by the standard deviation of the local continuum. In the top panel of Figure \ref{fig:lines}, we identify six lines that meet the $5\sigma$ criterion, with their wavelength positions automatically plotted above them. This function also provides options to set minimum line width and minimum separation, reducing the likelihood of spurious detections (see the GitHub tutorial for details).

At this stage, some expertise in astrophysical spectra interpretation is required. Users must be able to manually identify at least one spectral line before proceeding. In the case of AGN spectroscopy, the group of three emission lines around $8000\AA$ is readily recognized as H$\beta$ and two [OIII] lines. While identifying more lines improves accuracy, a single confirmed line is sufficient to continue the analysis. Next, we use the previously detected lines as input for the \texttt{analysis.fit\_lines()} function, which performs an MCMC estimation--based on the \texttt{emcee} Python library \citep{foreman2013emcee}--to retrieve the best-fit parameters for all lines, including both isolated and blended cases. This function supports parallelization and allows users to define priors for each line. If no priors are provided, easyspec selects them automatically as follows: (i) the line amplitude is initially set as the measured height of the line relative to the continuum--retrieved with the function \texttt{analysis.find\_lines()}--and allowed to vary between 0.1 and 10 times this value; (ii) the line center (wavelength position) is initially set as the observed wavelength of the line--also retrieved with the function \texttt{analysis.find\_lines()}--and allowed to vary within $\pm 100 ~\rm{\AA}$ for isolated lines, or within half the distance to the nearest neighboring line if it lies closer than $200 ~\rm{\AA}$; and (iii) the line width is initially set to $10 \rm{\AA}$ and allowed to vary in the range 0.1 to $150~\rm{\AA}$. Additionally, easyspec can be configured to display commonly observed elemental lines in astrophysical spectra. Its line database primarily consists of entries from the NIST Atomic Spectra Database \citep{ralchenko2005nist}\footnote{Most of the astrophysical lines used in easyspec are available here: \url{https://astronomy.nmsu.edu/drewski/tableofemissionlines.html}}\footnote{The NIST database can be found here: \url{https://www.nist.gov/pml/atomic-spectra-database}.}. For this specific case, we request Hydrogen, Oxygen, and Neon. This feature aids in the precise identification of spectral lines, even those below the $5\sigma$ threshold. Finally, the MCMC estimation provides a redshift measurement for the detected lines, as shown in the bottom panel of Figure \ref{fig:lines}. 

We then recover the highest likelihood parameters and corresponding asymmetrical errors (68\% confidence intervals based on the 0.16 and 0.84 quantiles of the posterior distributions) for all six fitted lines. The user can model all the lines with the same model, which can be Gaussian, Lorentzian, or Voigt, or choose a specific model for each line. Here we are modeling the H$\beta$ line with a Lorentzian, and all other lines with a Gaussian. To go deeper into the MCMC adopted in easyspec, including the corner plots with the covariances and temporal evolution of parameters, we refer the reader to the GitHub tutorials.


\section{Results}
\label{sec:results}

\begin{figure}
    \centering
    \includegraphics[scale=0.7]{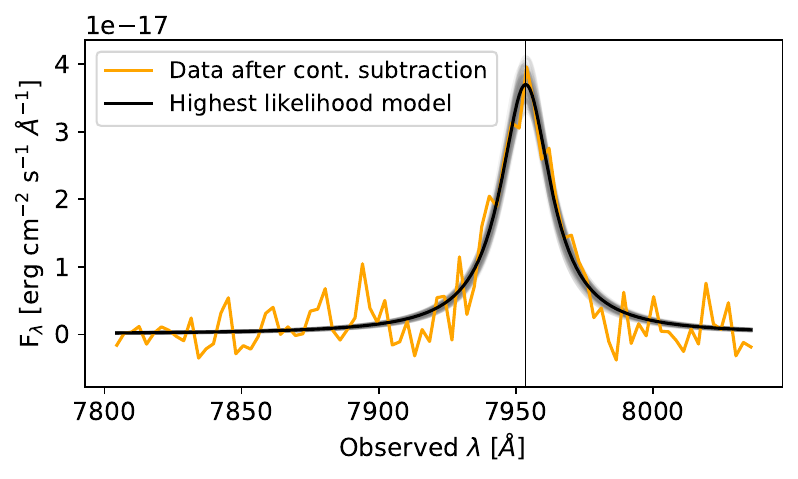}
    \caption{Lorentzian profile fit to the H$\beta$ line with the MCMC method. The fit is done in the observed frame, although the results are available in both the observed and rest frames.}
    \label{fig:Hbeta_line}
\end{figure}

Although we measured the parameters of 6 emission lines in the spectrum of the AGN G4Jy 1709 and found an average redshift of $z_{av} = 0.636158 \pm 0.000058$ (see bottom panel in Figure \ref{fig:lines}), from now on we focus only on the H$\beta$ line. With this line we can estimate the supermassive black hole mass with the scaling relationships described by \cite{vestergaard2006determining}. These formulas are available in easyspec in the \texttt{analysis.BH\_mass\_Hbeta\_VP2006()} function and can be explicitly written as follows:

\begin{equation}
    \label{eq:mass1}
    \log M_{BH}(H\beta) = \log \left[ \left(\frac{\rm{FWHM}(H\beta)}{1000~\rm{km/s}}\right)^2 \left(\frac{\lambda L_{\lambda}(5100~\AA)}{10^{44}~\rm{erg/s}}\right)^{0.5} + (6.91 \pm 0.02) \right],
\end{equation}
and
\begin{equation}
    \label{eq:mass2}
    \log M_{BH}(H\beta) = \log \left[ \left(\frac{\rm{FWHM}(H\beta)}{1000~\rm{km/s}}\right)^2 \left(\frac{L(H\beta)}{10^{42}~\rm{erg/s}}\right)^{0.63} + (6.67 \pm 0.03) \right],
\end{equation}
where FWHM stands for the full width at half maximum of the line, $\lambda L_{\lambda}(5100~\rm{\AA})$ is the rest-frame continuum luminosity at $5100~\rm{\AA}$, and $L(H\beta)$ is the line luminosity. The parameters of the Lorentzian fit performed in the $H\beta$ line (see Figure \ref{fig:Hbeta_line}) are listed in Table \ref{tab:H_beta_parameters}. By feeding these values to the \texttt{analysis.line\_physics()} function, we recover the line integrated flux $f_{H\beta} = 1.20^{+0.03}_{-0.03} \times 10^{-15}$ erg/s/cm$^2$--computed as the line equivalent width times the continuum value at the line center--and the rest-frame FWHM (already corrected for instrumental broadening, i.e., $FWHM = FWHM_{obs} - FWHM_{inst}$, where $FWHM_{inst} \approx \Delta\lambda$ at the position of the H$\beta$ line) in terms of the dispersion velocity $FWHM_v = 881^{+12}_{-10}$ km/s, where the conversion from $\rm{\AA}$ to km/s is done with the formula $FWHM_v = c \times FWHM /\lambda_0$, where $c$ is the speed of light and $\lambda_0$ is the line rest-frame wavelength. 

\begin{table}
    \centering
    \begin{tabular}{l|c|c}
        Parameter & value & unit\\
        \hline
        Redshift & $0.63610^{+0.00012}_{-0.00011}$ & -- \\
        Mean$_{obs}$ & $7953.61^{+0.53}_{-0.58}$ & $\AA$ \\
        Amplitude$_{obs}$ & $(3.67^{+0.18}_{-0.17})\times 10^{-17}$ & erg cm$^{-2}$ s$^{-1}$ $\AA^{-1}$ \\
        FWHM$_{obs}$ & $22.2^{+1.5}_{-1.3}$ & $\AA$
    \end{tabular}
    \caption{H$\beta$ line parameters recovered with the MCMC method. The fit is done in the observed frame, although easyspec converts everything to the rest-frame when needed. The amplitude is computed with respect to the local continuum, as shown in Figure \ref{fig:Hbeta_line}, while the mean represents the line peak position in the wavelength axis.}
    \label{tab:H_beta_parameters}
\end{table}

We now compute the rest-frame continuum luminosity at $5100~\rm{\AA}$ and the line luminosity, using the line flux and redshift obtained earlier, that is, $L(H_{beta}) = f_{H\beta} 4\pi D_L^2$, where $D_L$ is the luminosity distance. These values are then applied to Equations \ref{eq:mass1} and \ref{eq:mass2} to estimate the mass of the supermassive black hole, assuming a cosmological model with $H_0 = 70$ km/s/Mpc, $\Omega_{\Lambda} = 0.7$, and $\Omega_m = 0.3$. The resulting mass estimates, in solar masses, are $\log M_{BH} = 6.97 \pm 0.43$ and $\log M_{BH} = 6.71 \pm 0.43$, respectively, showing excellent agreement between the two methods.

This type of mass estimation is most reliable for AGNs with highly inclined accretion disks (i.e., close to $90^{\circ}$ relative to the observer's line of sight). In such cases, the full width at half maximum (FWHM) of emission lines provides an accurate measure of the total Doppler broadening due to the gas dynamics around the black hole, rather than just a projected component of this motion. For smaller inclination angles, a correction factor $f$ is needed to adjust the black hole mass estimate \citep[see e.g.,][]{marziani2012estimating}.

For G4Jy 1709, we have significant indicators suggesting a high inclination angle:
\begin{itemize}
    \item Despite being a powerful radio source \citep{massaro2023powerful}, it lacks detection in $\gamma$-rays by the Fermi Large Area Telescope \citep{abdollahi2020_4FGL, abdollahi2022_4FGL_DR3}.
    \item The equivalent width of its [O III]$\lambda 5007~\rm{\AA}$ line, which is proposed as an orientation indicator \citep{risaliti2011iii,bisogni2017orientation}, is relatively large (i.e., $EW = -221.29^{+1.10}_{-0.94}\rm{\AA}$), suggesting that our target is viewed from a high inclination angle.
\end{itemize}

Given these factors, it is reasonable to assume a correction factor of $f \approx 1$, indicating that no additional correction to the black hole mass estimate is necessary. Additionally, due to the well-established anti-correlation between Fe II and [O III] emission in quasar spectra \citep{boroson1992emission}, we do not expect a significant iron contribution in this spectrum. Moreover, since G4Jy 1709 is not a blazar, its continuum emission is unlikely to be contaminated by non-thermal emission from AGN jets. Therefore, no subtraction of non-thermal emission is required \citep[see, e.g.,][]{raiteri2019beamed, raiteri2020unveiling}.

\section{Validation}
\label{sec:validation}

\begin{figure*}
    \centering
    \includegraphics[width=\linewidth]{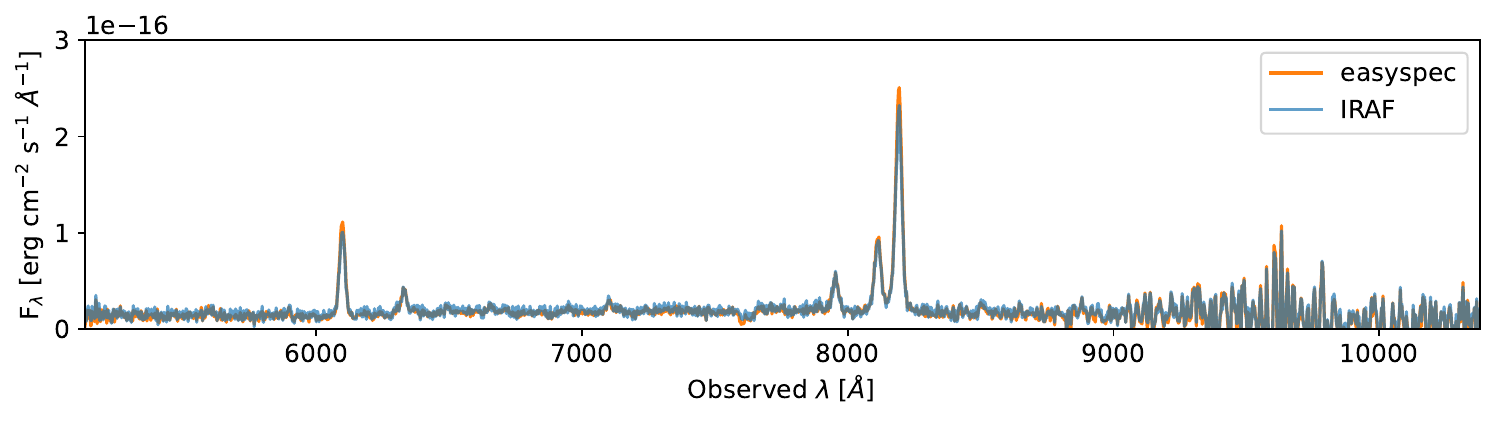}
    \caption{Comparison between the spectral data reduction performed with easyspec and IRAF, showing an excellent agreement between the two data reductions.}
    \label{fig:IRAF}
\end{figure*}

To validate our results, we perform the same spectroscopic analysis with IRAF, following the standard routines as done in \cite{deMenezes2020optical_camp_X}. The comparison between the two data reduction methods is shown in Figure \ref{fig:IRAF}. While minor differences can be observed, the overall spectra are nearly identical.

We can furthermore compare the average redshift $z_{av} = 0.636158 \pm 0.000058$ we measured for G4Jy 1709 with that reported in \cite{holt2008fast}, $z_{Holt} = 0.63634 \pm 0.00003$. In their study, the authors used IRAF to process a 2D spectrum obtained with the William Herschel Telescope in La Palma, Spain. The excellent agreement between these measurements--consistent within a 1/10000 fraction--further supports the reliability of our reduction process.

Another key comparison with \cite{holt2008fast} is the FWHM of the [OIII]$\lambda 5007 \rm{\AA}$ emission line. Our MCMC analysis (Sect. \ref{sec:line_fitting}) yields $FWHM_v = 881^{+12}_{-10}$ km/s, while in \cite{holt2008fast} the reported value is $FWHM_{Holt} = 919 \pm 7$ km/s. Here we see a $\sim 4\%$ discrepancy between the results, which are still compatible within $2\sigma$ error bars. We take this as a decent level of agreement given that their observations were conducted over 15 years prior using a different instrument. Possible intrinsic spectral variations or systematic differences in data reduction could account for this small offset.

\section{Discussion and conclusions}
\label{sec:discussion_and_conclusions}

In this work, we introduced easyspec, a Python package designed for the analysis of near-infrared, optical, and near-ultraviolet long-slit spectra. As a demonstration of its capabilities, we applied it to the AGN G4Jy 1709, measuring its redshift as $z_{av} = 0.636158 \pm 0.000058$--a value fully consistent with previous literature. 

Moreover, we provided the first estimation of its supermassive black hole mass, obtaining $\log M_{BH} = 6.97 \pm 0.43$ and $\log M_{BH} = 6.71 \pm 0.43$ (in terms of solar masses) using two independent scaling relations. These results highlight easyspec as a powerful, flexible, and user-friendly tool for long-slit spectroscopy data reduction and analysis.

The applications of easyspec are extensive, spanning various fields of astronomy, including the study of stars, nebulae, and galaxies. The software is compatible with instruments that provide raw FITS images, making it a versatile tool for spectroscopic data analysis. So far, we have successfully tested easyspec on data from multiple telescopes, such as TNG (as shown in the previous sections), the 4.1m SOAR telescope, in Chile, and the 1.6m Perkin-Elmer telescope at the OPD, Brazil. The development of easyspec is part of a broader effort to create modern Python-based tools for astronomers. It follows in the footsteps of its predecessor, easyfermi \citep{deMenezes2022easyfermi}\footnote{\url{https://github.com/ranieremenezes/easyfermi}}, and is the second in a series of open-source packages planned to facilitate astronomical data analysis.

The potential of easyspec extends far beyond the analysis presented here. The software allows users to:
\begin{itemize}
    \item Fit multiple components (Gaussian, Lorentzian, or Voigt) to a single spectral line.
    \item Monitor the temporal evolution and covariances of MCMC parameters.
    \item Detect spectral absorption lines.
    \item Compute equivalent widths and velocity dispersions of spectral lines.
    \item Fit blended lines (up to three overlapping lines).
    \item Perform bulk data cleaning for several targets at once (provided they share suitable bias and flats).
    \item Extract multiple 1D spectra from a single 2D spectral image.
    \item Vertically align exposures in case the telescope loses track between observations.
\end{itemize}
Additionally, easyspec is highly customizable and can be integrated into complex pipelines, depending on the user's needs. All of these features are described in detail in the GitHub tutorials and documentation. Another noteworthy capability is that the \texttt{cleaning()} class can also be applied to photometric data reduction, further expanding the software's versatility.

\begin{acknowledgments}
We thank the anonymous referee for the detailed report and for the constructive comments that allowed us to substantially improve the manuscript. C.M.R. acknowledges financial support from the INAF Fundamental Research Funding Call 2023. J.A.-P. acknowledges financial support from the Spanish Ministry of Science and Innovation (MICINN) through the Spanish State Research Agency, under Severo Ochoa Centres of Excellence Programme 2020–2024 (CEX2019-000920-S).
\end{acknowledgments}

\begin{contribution}

R.M. was responsible for developing the code and drafting the manuscript. F.M. and M.N. provided the optical spectra featured in the main text and the GitHub tutorials, as well as several conceptual contributions that were incorporated into the code. C.M.R. contributed the theoretical framework for measuring black hole masses and offered numerous suggestions for the spectroscopy reduction process. H.P.H. and J.A.A.P. provided valuable insights regarding the code's features and the data reduction procedures. All authors participated in the critical revision of the manuscript.

The initial idea for this work originated when F.M. approached R.M. with a request to assist in teaching a group of students how to use legacy software for optical spectra reduction. R.M.'s decision to decline this request ultimately led to the creation of easyspec.


\end{contribution}

%
\facilities{TNG(DOLORES), SOAR(Goodman)}



\bibliography{sample7}{}
\bibliographystyle{aasjournalv7}



\end{document}